\documentclass{JHEP3}
\title{Quantum broadening of  k-strings in gauge theories}
\author{Pietro Giudice, Ferdinando Gliozzi, Stefano Lottini\\
Dipartimento di Fisica Teorica, Universit\`a di Torino and \\
INFN, Sezione di Torino\\
via P.Giuria 1, I-10125 Torino, Italy\\
\email{giudice,gliozzi,lottini@to.infn.it}}
\abstract{We study the thickness of the confining flux tube generated by 
a pair of sources in higher representations of the gauge group. Using a simple 
geometric picture we argue 
that the area of the cross-section of the flux tube, as measured by a 
Wilson loop probe, grows logarithmically with source separation, as a 
consequence of the quantum fluctuations of the underlying k-string. 
The slope of the logarithm turns out to be universal, i.e. it is the same for 
all the representations and all the gauge theories. We check these predictions 
in a 3D $\Z_4$ lattice gauge model by comparing the broadening of the 1-string
and the 2-string.}

\usepackage[dvips]{graphicx}
\usepackage{amsmath}
\usepackage{amsfonts}
\usepackage{mathrsfs}
\usepackage{epsfig}
\usepackage{psfrag}
\newcommand{\Z}{\mathbb{Z}}

\newcommand{\R}{\mathscr{R}}

\newcommand{\bra}{\langle}
\newcommand{\ket}{\rangle}
\newcommand{\avg}[1]{\langle \hspace{0.2em} #1 \hspace{0.2em} \rangle}

\newcommand{\sun}{\mathop{\rm SU}(N)}

\newcommand{\eq}{\begin{equation}}
\newcommand{\en}{\end{equation}}
\newcommand{\bea}{\begin{eqnarray}}
\newcommand{\ea}{\end{eqnarray}}
\newcommand{\arccosh}{{\rm arccosh}}
\newcommand{\link}[1]{\bra #1\ket}
\newcommand{\overtilde}[1]{\stackrel{\sim}{#1}}
\begin{document}
\section{Introduction}
In the physics of quark confinement there are many indications that the gauge  
field responds to a static $q$ source  separated from  a 
conjugate  $\bar q$ source by a large distance $L$ by forming a colour 
flux tube which behaves as a string with energy $V(L)\simeq \sigma_{\R}\,L$,
where the string tension $\sigma_\R$ depends on the representation $\R$ 
of the quantum numbers carried by the source. 

If the gauge group is $\sun$ there are infinitely many irreducible 
representations at our disposal
to cast the sources. However, for large separations, no matter what 
representation is chosen, $\sigma_{\R}$  depends only on the   
$N-$ality $k$ of $\R$, i.e. on the number (modulo $N$) of copies of the 
fundamental representation needed to build $\R$ by tensor product, the reason being that 
all representations with the same $k$ can be transformed into each other 
by the emission of a proper number of soft gluons. As a consequence the 
heavier strings decay into the string of smallest string tension. The 
corresponding string is referred to as a k-string.

The spectrum of k-string tensions has been extensively studied in 
recent years, in the continuum \cite{ds}--\cite{Armoni:2006ri} as well as  
on the 
lattice \cite{lt1}--\cite{ddpv}. In this paper we want to explore another 
facet of k-string physics, related to the quantum fluctuations of these 
objects. 

It is widespread belief  that the flux tube generated by a pair of sources 
$q$, $\bar q$ in the fundamental representation  
is in  the rough phase. This means, as explained long ago by 
L\"uscher, M\"unster and Weisz \cite{lmw}, that the colour flux tube  
broadens as the sources are separated. More precisely, the area of a 
cross-section of the flux tube should increase logarithmically with 
separation. This phenomenon can be seen  as a consequence of the 
quantum vibrations of the  underlying 1-string 
describing of the infrared properties of this flux tube. 

Since the k-strings can be seen as bound states of $k$  1-strings, 
it is interesting to ask if there is any fundamental obstruction
for a logarithmic broadening of the k-strings.
At first sight it is not clear whether in the IR 
limit the only relevant degrees of freedom are the transverse displacements
of a single string or whether we have to take into account new degrees of 
freedom describing the possible splitting of the $k$-string into its 
constituent strings. In other terms one would 
like to know  whether the string binding energy  damps 
significantly  the quantum fluctuations of the free strings.

The answer we find to the above questions is surprisingly simple: not only the 
logarithmic broadening occurs  in the k-strings for any $k$, but it turns also 
out that the coefficient of the logarithm does not depend on 
the specific representation of the source nor on its N-ality and is 
universal, i.e. it is the same in all gauge theories.
We check these predictions in a 3D $\Z_4$ gauge model, which is the simplest 
environment where a non-trivial 2-string can live. A brief report of our work 
has been presented in \cite{Giudice:2006xe}.

Measuring the thickness of the flux tube and in particular its dependence on 
the separation of the sources  is very 
challenging from a computational point of view. In $\sun$ gauge theories
the error bars are too large to draw definite conclusions \cite{ba}.
As a matter of fact, 
an uncontroversial observation of logarithmic broadening has been 
only made in 3D $\Z_2$ gauge model, thanks to the efficiency of the 
Monte Carlo algorithms for its dual, the Ising model \cite{cgmv}.
New results on the thickness of the ${\rm SU}(2)$ flux tube near the 
deconfining point exploiting the integrability of the underlying 2D 
Ising model appeared recently \cite{Caselle:2006wr}. 
The numerical data 
on the logarithmic growth of the mean squared width of the flux tube 
associated to the $\Z_4$ strings presented here constitute
a new  important support on the expected quantum behaviour of the flux tube.  

\section{The rough phase}
In the strong coupling phase of whatever gauge theory in three or four 
space-time dimensions the flux tube joining a quark pair has a 
constant width for large inter-quark distances. As the coupling constant 
decreases, the flux tube can undergo a roughening transition. The rough 
phase is 
characterised by strong fluctuations of the collective coordinates 
describing the position of the underlying string and the mean squared width 
of the flux tube diverges logarithmically when the inter-quark distance goes 
to infinity \cite{lmw}.

The square width of the flux tube generated by  a planar Wilson loop $W_f(C)$
in the fundamental representation is defined as the sum of the mean square 
deviations of the transverse coordinates $h_j(\xi_1,\xi_2)$ $(j=1,\dots D-2)$ 
of the underlying string, i.e.
\eq
w^2=\frac1A\sum_{j=1}^{D-2}\int_\Omega  \mathrm{d}^2\xi\bra
(h_j(\xi_1,\xi_2)-h_j^o)^2\ket\,,
\en
 where $\Omega$ is the planar domain bounded by $\partial\,\Omega=C$, 
$A=\int_\Omega  \mathrm{d}^2\xi$ its area and $h^o_j$ are the transverse 
coordinates of the equilibrium position. The vacuum expectation value is 
taken with respect to the two-dimensional field theory describing the 
dynamics of the flux tube. It is widely believed that the roughening 
transition belongs to the Kosterlitz-Thouless universality class \cite{kt}. 
As a consequence, it is expected that this field theory flows, 
in the infrared (IR) limit,  toward the massless free-field theory 
described by the Gaussian action
\eq
S_{IR}=\frac\sigma2\sum_{j=1}^{D-2}\int_\Omega  
\mathrm{d}^2\xi \sum_{\mu=1,2}\partial_\mu h_j  \partial^\mu h_j\,.
\label{irl}
\en 
 In such a limit the mean square width can be easily evaluated in terms of the
free Green functions $G_\Omega(\xi,\xi')=\bra h(\xi)\,h(\xi')\ket$ as
\eq
w^2=\frac{D-2}{A^2}\int_\Omega  \mathrm{d}^2\xi \;\int_\Omega  
\mathrm{d}^2\xi'\left(G_\Omega(\xi,\xi')-G_\Omega(\xi,\xi+\epsilon)\right)\,,
\label{intg}
\en
where $\epsilon$ is a UV cut-off. The action (\ref{irl}) is conformally  
invariant, hence the integration of the finite part $G(\xi,\xi')$
cannot depend on the size of the domain $\Omega$ but only on its shape. 
The logarithmic growth of $w^2$ comes from the UV divergent part.
This can be simply understood as follows \cite{cgmv}.
The conformal invariance of the theory implies the scaling property
\eq
G_\Omega(\xi,\xi')=G_{\Omega_s}(s\,\xi,s\,\xi')\,,
\label{scaling}
\en 
where $s>0$ is an arbitrary real number and $\Omega_s$ is the scaled domain. 
On the other hand in the UV limit $\xi'\to\xi$ the Green function diverges 
logarithmically
\eq
G_\Omega(\xi,\xi+\epsilon)=-\frac1{2\pi\sigma}\log\epsilon+ \dots\,.
\label{diverge}
\en
(\ref{scaling}) and (\ref{diverge}) agree only if the cut-off appears 
exclusively 
in the ratio $\epsilon/R$, where $R$ is a typical linear size of the domain.
Adding this piece of information to (\ref{intg}) yields
\eq
w^2=\frac{D-2}{2\pi\sigma}\log(R/R_f(\Omega))\,,
\label{msw}
\en
where the UV cut-off has been absorbed in the scale $R_f(\Omega)$, thus the 
absolute value of this physical scale cannot be determined by this conformal
approach. On the contrary  the  ratios of these scales for different domains 
are calculable functions of the shapes \cite{cgmv,kp}.

It is clear from this approach that the generalisation to the k-string 
crucially depends on its IR limit. If, for instance, it 
fluctuated as a single string it would suffice to put $\sigma\to\sigma_k$
in (\ref{irl}) and modify the other formulae consequently. There are  
other possibilities, however. We shall see in the next section that in $D=3$ 
dimensions there is a 
geometric approach, first advocated in \cite{lmw}, which can be unambiguously
extended to the k-strings.   

\section{K-strings and minimal surfaces}
To inspect the width of the flux tube generated, in a 3D gauge system, by a static source in the 
 fundamental representation one can start, following 
 \cite{lmw}, 
from the connected correlator
\eq
P_f(h) = \frac{ \avg{W_f(C)\,W_f(c)} - 
\avg{W_f(C)}\avg{W_f(c)}}{\avg{W_f(C)}}\,,
\label{pfh}
\en
where $W_f(C)$ and $W_f(c)$ are Wilson operators for the loops $C$ and $c$, 
which are 
concentric circles of radii $R$ and $r < R$ lying on parallel planes at a 
distance $h$, as shown in Fig.~\ref{Figure:1}. The above quantity 
can be seen as a measure of the  
density of the colour flux generated by 
the source $C$  and felt by the probe $c$. 
\FIGURE{\epsfig{file=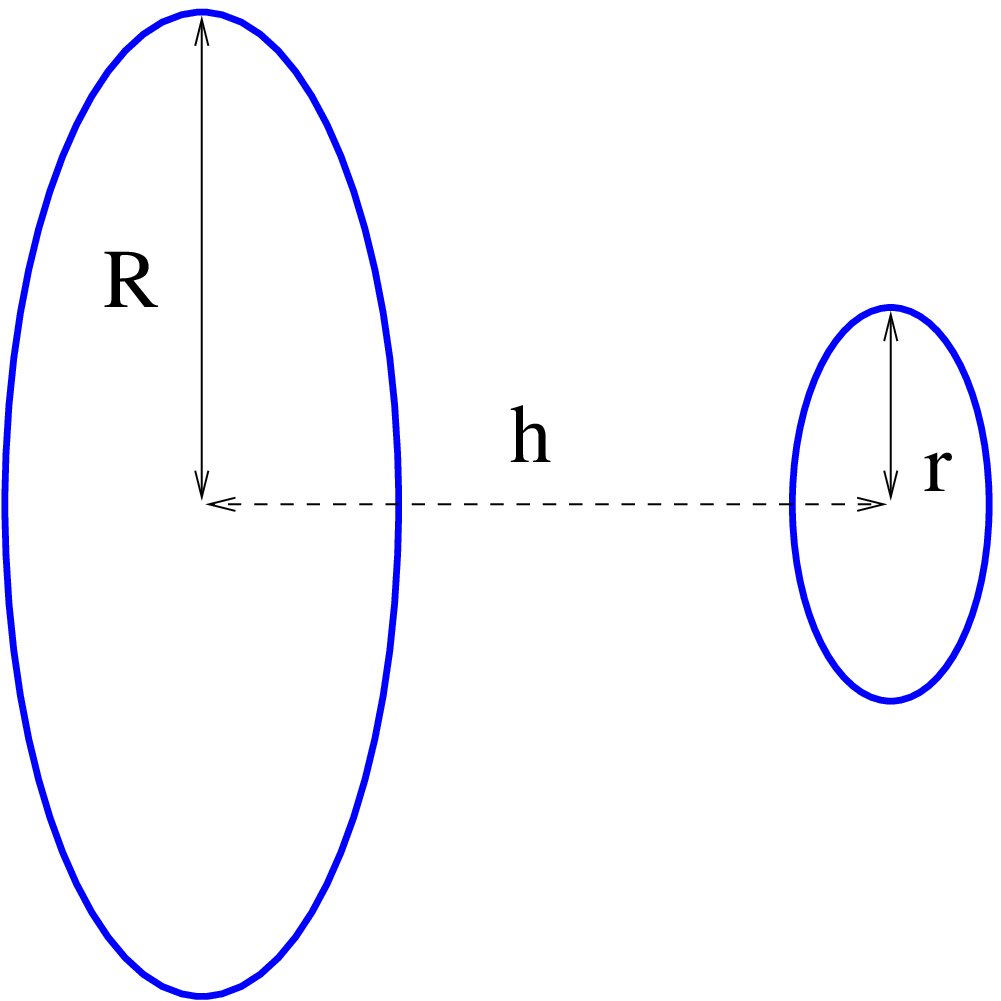,width=5 cm}
\caption{Parallel circular loops \label{Figure:1}}}
To study the flux generated by sources in a generic representation 
$\mathscr{R}$, we will take $W_{\mathscr{R}}(C)$ instead of $W_f(C)$, but 
keep the probe always in the fundamental.

The mean square width of the flux tube is defined by
\eq
	w^2_{\mathscr{R}}=\frac{\int h^2 P_{\mathscr{R}}(h)\,
\mathrm{d}h}{\int P_{\mathscr{R}}(h)\, \mathrm{d}h}\,.
\label{eq:defw2}
\en

The insight of \cite{lmw} was to observe that the quantity $P_f(h)$ 
can be described in 
the effective string picture in terms of the world-sheet of a 
Nambu-Goto string  connecting 
the two circles. This defines  a typical Plateau problem of minimal surfaces, 
which can  be evaluated in the infrared limit by a saddle-point 
approximation. 
More specifically, one can write
\eq
	\avg{W_f(C)\,W_f(c)} - \avg{W_f(C)}\avg{W_f(c)} 
\propto \exp(-\sigma A(R,r,h))\,,
\en
where  $A(R,r,h)$ denotes the area of the connected minimal
 surface having $C$ and $c$ as boundaries. 
\FIGURE{\epsfig{file=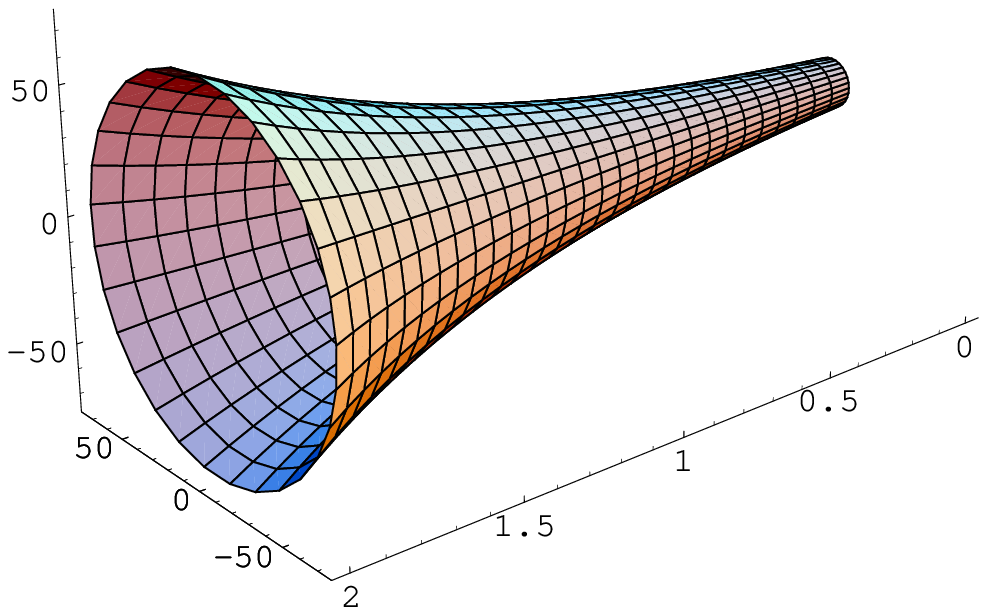,width=7 cm}\caption{Catenoid surface
\label{Figure:2}}}
The choice of the Nambu-Goto (NG) action for the effective string is the 
simplest one, however there are well known problems for its quantisation 
procedure. This theory is fully consistent only  at critical space-time 
dimensions (D=26)\footnote{Polchinski and Strominger \cite{ps} developed an effective 
string theory which avoids quantisation problems in non-critical physical 
dimensions.}, but there is strong evidence that the first few terms in 
the $1/(\sigma\, A)$ expansion of the NG action are  universal 
up to the order $O \left[\frac1{(\sigma\, A)^2}\right]$  
(see \cite{jk}, and references quoted therein, for a detailed discussion on this point). In particular, the $O(1)$ term coincides with (\ref{irl}), thus it is 
reasonable to expect that in the infrared limit $R\to \infty$ with fixed $r$ 
and $h$ the result should not depend on the specific choice of NG action.
 Note however that the size $r$ of the probe cannot be too small, otherwise 
the string picture would not be valid.

The minimal surface connecting the two circles is a surface of revolution 
about the 
symmetry axis (here chosen to coincide with the $\boldsymbol{x}$ axis). 
If we denote by $y(x)$ the $y>0$ section of the surface with the $(x,y)$ 
plane, the area is given by
\eq
	A = 2\pi \int_0^h y \sqrt{\dot{y}^2+1}\, \mathrm{d} x\,.
\en

The variational condition $\frac{\delta A}{\delta y(x)}=0$ yields the 
equation  $1+\dot{y}^2 = y\, \ddot{y}$. The general solution is
\eq
	y(x) = \frac{1}{\omega} \cosh\omega(x-x_0)\,,
\en
that is, the surface of revolution is a catenoid (see Fig. \ref{Figure:2}). 
The integration constants must obey
\eq
	R = \frac{1}{\omega}\cosh\omega x_0 \quad , 
\quad r=\frac{1}{\omega}\cosh\omega(h-x_0)\,.
\label{Rr}
\en
This results in the following expression for the minimal area:
\eq
	A = \pi\Big(\frac{h}{\omega}+R^2\sqrt{1-\frac{1}{\omega^2R^2}}-
r^2\sqrt{1-\frac{1}{\omega^2r^2}}\Big)\,;
\en
moreover Eq. (\ref{Rr}) allows to express  $h$ as a decreasing  
function of $\omega$ 
\eq
	h(\omega)=\frac{\big[\arccosh(R\omega)-\arccosh(r\omega)
\big]}{\omega}\,,
\label{eq:accadiomega}
\en
thus we can regard $\omega$ as an integration variable for actually 
computing $w^2$. 
Using the trivial inequality  $\cosh x\geq 1~(\forall\,x)$ we get a 
minimal value for $\omega$, that is a maximal allowed value for $h$:
\eq
	\omega_{min} = \frac{1}{r} \quad \Rightarrow \quad h_{max} = 
h(1/r) \,.
\en
Now we can write explicitly, for the mean square width,
\eq
	w^2=\frac{\int_{\frac{1}{r}}^\infty h(\omega)^2\,
exp[-\sigma A(R,r,h(\omega))]\,
	\vert h'(\omega)\vert\,d\omega}{
	\int_{\frac{1}{r}}^\infty \,exp[-\sigma A(R,r,h(\omega))]\,
	\vert h'(\omega)\vert\,d\omega}\,.
\en

This quantity approaches a logarithmic curve for large $R$. Indeed, 
this can be seen by using the asymptotic expansion
\eq 
\arccosh(\omega r) \sim 2 \log(\omega r)\;\;\;\;(\omega r\to\infty)\,,
\en 
and inserting it into (\ref{eq:accadiomega}), 
which yields the approximate solution found in \cite{lmw}:
\eq
	\omega \sim \frac{1}{h}\log(R/r)\,.
\en

The condition $\omega r \gg 1$ becomes 
\eq
\log (R/r) \gg h/r\,,
\en
always fulfilled in the large $R$ limit. Note that  $r$ cannot be too 
small at fixed $R$ and $h$. 
In this limit, a Gaussian distribution is found 
for the transversal density, whose width grows logarithmically with $R$:
\eq
	P^{(R)}(h) \propto 
\exp\left[-\frac{\sigma \pi h^2}{\log(R/r)}\right] \quad \Rightarrow \quad
               \sigma w^2 = \frac{1}{2 \pi} \log (R/r)\,,
\label{logr}
\en
which is almost identical to (\ref{msw}). Here the UV cut-off is replaced 
by the size $r$ of the probe.  

The result found here can be generalised to Wilson loop 
$W_\mathscr{R}(C)$ in any representation (where we are no 
more guaranteed that the infrared Gaussian free string limit is valid). 
The world-sheet of the k-string can be seen as some bound state of $k$ 
1-string world-sheets. In this case a local minimum solution exists 
for sure, in which one of these fundamental sheets gives rise to the 
catenoid with the probe $W_f(c)$, while the other $k-1$ just lie flat on 
the loop surface. We then have immediately
\eq
	P^{(R)}_{(k)}(h) = P^{(R)}_{(1)}(h) \cdot \exp[\pi R^2 
(\sigma_k - \sigma_{k-1} - \sigma)]\,,
\en
where the stability of the k-string implies that the exponent is 
always negative. The resulting width, then, appears to be exactly 
the same as for the fundamental representation.

\subsection{Special configurations}
In some special cases, for a very narrow range of $h$, there is, beside 
the above general solution, another minimal surface made by a first 
catenoid composed by the k-string world-sheet.
\FIGURE{
\psfrag{a}{$\theta_1$}
\psfrag{b}{$\theta_2$}
\psfrag{c}{$\theta_3$}
\psfrag{T2}{$\sigma_{k_2}$}
\psfrag{T1}{$\sigma_{k_1}$}
\psfrag{T3}{$\sigma_{k_3}$}
\epsfig{file=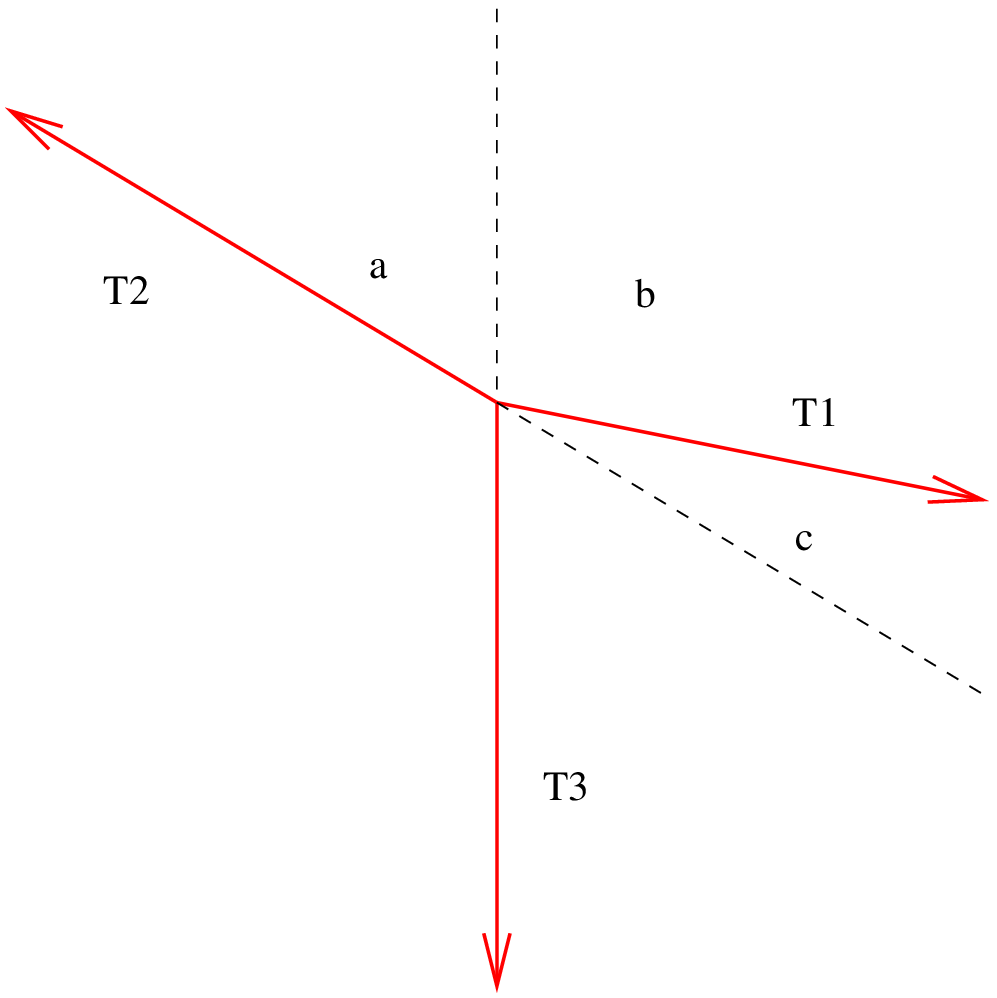,width=4 cm}
\caption{The balance of the string tensions\label{Figure:3}}}
At a suitable distance $d$ it splits into a disk orthogonal to 
the symmetry axis made by the world-sheet of the $(k-1)$-string and a 
second catenoid in the fundamental representation which reaches the probe $c$. 
The position of the intermediate  disk is not arbitrary, but it is 
dynamically determined by the balance of 
the tensions at the string junction. 
In the most general configuration depicted in 
  Fig. \ref{Figure:3} a $k_2$-string propagating along a 
catenoid decays into a $k_3$-string forming a disk and  a $k_1$-string 
forming another catenoid. The angles of the string junction are given by

\eq
	\cos \theta_1 = \frac{\sigma_{k_2}^2+
\sigma_{k_3}^2-\sigma_{k_1}^2}{2\sigma_{k_2}\sigma_{k_3}}\,,
\label{costh1}
\en
and  cyclic permutations of the indices for the other angles.

Implementing these kinematic constraints on the parameters $\omega_0,x_0$ 
and $\omega_1,x_1$ of the two catenoids yields
\bea
\omega_0\,R=&\cosh\omega_0\,x_0~,\,\omega_0\,\rho=
\cosh\omega_0(x_0-d)=1/\sin\theta_1\,,\\
\omega_1\,r=&\cosh\omega_1(x_1+d-h)~,\,\omega_1\,\rho=
\cosh\omega_1x_1=1/\sin\theta_2\,,
\ea
where $d$ is the distance of the disk from the source and $\rho$ its radius. 
We can invert these relations and express $d$ and $h$ as functions of $\rho$
\bea
d(\rho)=&\rho\,\sin\theta_1\,\arccosh(R/\rho\,\sin\theta_1)-\rho\,
\sin\theta_1\,\arccosh(1/\sin\theta_1)~,\\
h(\rho)=&\rho\,\sin\theta_2\,\arccosh(1/\sin\theta_2)-\rho\,\sin\theta_2\,
\arccosh(r/\rho\,\sin\theta_2)+d(\rho)\,.
\ea
Both $d(\rho)$ and $h(\rho)$ turn out to be increasing functions of $\rho$.
Since $h\ge d $ we have $\rho\ge r$ which implies $\theta_2\ge \frac\pi2 $, 
as Fig. \ref{Figure:3} shows, then (\ref{costh1}) tells us that this special surface
 is permitted only if  the kinematic constraint 
\eq
\sigma_{k_2}^2 \ge \sigma^2_{k_1}+\sigma^2_{k_3}
\label{constraint}
\en 
holds, i.e. only when the binding energy of the $k$-string is 
sufficiently small. On the other hand,
the fact that the argument of 
$\arccosh$ must be larger or equal to 1 fixes the maximum $\rho_{max}
=r/\sin\theta_2$.
Thus the minimal surface exists only in the range
\eq
h(r)\leq h(\rho)\leq h(r/\sin\theta_2)\,,
\label{hrange}
\en
which turns out to be in general very narrow. These special configurations 
should produce a spike in the distribution of the flux density at the distance
fixed by (\ref{hrange}).

The hight of such a peak is determined by the total area of the surface. 
Unfortunately this area  depends significantly on $r$, 
hence it is rather unclear how to extract  physical 
information on the flux tube, which should not depend 
on the size of the probe.  

In the $N=4$, $k=2$ case, which corresponds to our  simulations,
 this special solution should obey the constraint (\ref{constraint}) i.e. 
$\sigma_2\ge\sqrt{2}\,\sigma$.
In our numerical data taken in that region we did not find signs of
such a singular behaviour. 

\section{$\Z_4$ gauge theory and its dual}

The laboratory where we study the physical properties of the 2-string is 
a $\Z_4$ gauge model in three dimensions, defined by the standard plaquette 
action on a cubic lattice. With $\Z_4$ as gauge group, there exist two 
k-strings in the system: the fundamental string and the 2-string (related 
to the double-fundamental representation $f\otimes f$). Moreover, since 
there are no more representations than one in the same N-ality class, 
the system does not exhibit any meta-stable string that could spoil the 
results at finite $R$.

It is well known that this model, as any three-dimensional abelian 
gauge model, admits a spin model as its dual (see for instance \cite{id}) 
and any physical property of the gauge system can be translated into a 
corresponding property of its spin dual. From a computational point of view 
it is of course much more convenient to work directly on the spin model 
where powerful non-local cluster algorithms can be applied.

In our case the dual is a spin model with global $\Z_4$ symmetry which can 
be written as a symmetric Ashkin-Teller (AT) model \cite{at}, i.~e.~two 
coupled, ferromagnetic, Ising models defined by the two-parameter action
	\eq
	S_{AT} = -\sum_{\link{xy}} \Big[ \beta (\sigma_x \sigma_y + \tau_x \tau_y) 
                 + \alpha (\sigma_x \sigma_y \tau_x \tau_y) \Big]\,,
	\en
where $\sigma_x$ and $\tau_x$ are the Ising variables ($\sigma,\tau=\pm 1$) 
associated to the site $x$ and the sum is over all the links $\link{xy}$ of 
the dual cubic lattice. The phase diagram of this model has been studied long
 ago \cite{dbgk,az}.

The global $\Z_4$ symmetry of the action is generated by the transformation
\eq 
\sigma \to - \tau \: ; \: \tau \to \sigma\,.
\label{genz4} 
\en
The model has also an 
independent $\Z_2$ symmetry generated by
\eq 
\sigma\leftrightarrow\tau\,,
\label{chagen}
\en
 related to the charge conjugation of the corresponding dual model.

It is customary to represent the $\Z_4$ symmetry through the multiplication
 table of the fourth roots of the identity ($\xi_j = \pm i, \pm 1$). It would 
be a simple exercise to rewrite the AT model with this kind of variables 
introducing the complex field
\eq
 \Psi_x=e^{-i\frac{\pi}{4}}
(\sigma_x+i \tau_x)\,,
\label{psi}
\en
 where the phase factor is chosen such that 
the $\Z_2$ symmetry defined above becomes the complex conjugation $\Psi
 \leftrightarrow \Psi^\star$. The $\Z_4$ symmetry, instead, corresponds
 to the multiplication by $i$.

The standard application of the duality transformation would lead to a
 gauge model on the dual lattice with a $\{\xi_j\}$-valued field on the links 
and with an action containing plaquette operators and their squares:
	\eq
	S = -\sum_p \left[ c_1 \mathfrak{Re}(U_p) + c_2 (U_p)^2 \right] \,.
	\en
We think it is interesting (and more purposeful for our work) to show that 
the AT model is dually equivalent to \emph{two coupled} $\Z_2$ gauge models.

The starting point is to rewrite the Boltzmann factors associated to the 
links in the known form, namely the $\Z_2$ character expansion
\begin{eqnarray}
	e^{\beta\, \sigma_x \sigma_y} & = & \sqrt{\frac{\sinh 2 \beta}2}
 \sum_{\mu_{\link{xy}}=\pm 1} e^{\overtilde{\beta}\,\mu_{\langle xy\rangle}} 
(\sigma_x \sigma_y)^{\hat{\mu}_{\link{xy}}} \label{eq:char-1} \\
	e^{\beta\, \tau_x \tau_y} & = & \sqrt{\frac{\sinh 2 \beta}2} 
\sum_{\nu_{\link{xy}}=\pm 1} e^{\overtilde{\beta}\,\nu_{\link{xy}}} 
(\tau_x \tau_y)^{\hat{\nu}_{\link{xy}}} \\
	e^{\alpha\, \sigma_x \sigma_y \tau_x \tau_y} & = &
 \sqrt{\frac{\sinh 2 \alpha}2} \sum_{\rho_{\link{xy}}=\pm 1} 
e^{\overtilde{\alpha}\,\rho_{\link{xy}}}
 (\sigma_x \sigma_y \tau_x \tau_y)^{\hat{\rho}_{\link{xy}}}
\end{eqnarray}
with $\overtilde{\gamma} = -\log \sqrt{\tanh \gamma}$ and 
$\hat{\eta} = \frac{1}{2} - \frac{\eta}{2}$.

Now the sum over the site variables $\sigma$ and $\tau$ can be 
explicitly performed, yielding for each node $x$ the two conservation laws
\eq
	\sum_y \left( \hat{\mu}_{\link{xy}} + \hat{\rho}_{\link{xy}} \right)
\equiv 0 \: 
\textrm{mod } 2
	\: \textrm{,} \: \sum_y \left( \hat{\nu}_{\link{xy}} + 
\hat{\rho}_{\link{xy}} \right)
 \equiv 0 \: \textrm{mod } 2
\en
or, equivalently, in the multiplicative form
\eq
	\prod_y \mu_{\link{yx}}\rho_{\link{yx}} = 1~~~ \: \textrm{,} \: 
\prod_y \nu_{\link{yx}}\rho_{\link{yx}} = 1\,.
	\label{eq:constraints}
\en
The canonical partition function becomes
\eq
	Z_{AT} \propto \sum^{'}_{\mu=\pm 1, \nu=\pm 1, \rho=\pm 1}
 \exp{\sum_{\link{xy}} \left[ \overtilde{\beta}(\mu_{\link{xy}}+
\nu_{\link{xy}}) + 
\overtilde{\alpha} \rho_{\link{xy}}\right]} \,.
\en

The apex in the sum over configurations indicates that they must obey the
 constraints (\ref{eq:constraints}). A way to solve them in an infinite 
lattice is suggested by the usual duality transformation of the 
three-dimensional Ising model: it is sufficient to consider the two 
composed link variables $\mu_{\link{yx}}\rho_{\link{xy}}$ and 
$\nu_{\link{xy}}\rho_{\link{xy}}$ as the \underline{plaquette} 
variables of the dual lattice.
 More precisely we solve the above constraints with the Ansatz
\eq
	\mu_{\link{xy}} = \epsilon_{\link{xy}} U_P \: , \: \nu_{\link{xy}} = 
\epsilon_{\link{xy}}
 V_P \: , \: \rho_{\link{xy}} = \epsilon_{\link{xy}} U_P V_P \: , 
\label{eq:signum}
\en
here $\epsilon$ is an arbitrary sign variable ($\epsilon = \pm 1$), $P$ is
 the plaquette dual to the link $\link{xy}$. The plaquette variables are 
defined through the product of their boundary links, namely,
\eq
	U_P = \sum_{\ell \in \partial P} U_\ell \: , \: V_P = \sum_{\ell \in 
\partial P} V_\ell \: ,
\en
where $\ell$ are links of the dual lattice, of course. It is a straightforward 
exercise to verify that the above Ansatz solves identically Eq.s 
(\ref{eq:constraints}). The sum over the $\epsilon$ variables is 
unconstrained and can be performed at once, leading to
\eq
	Z_{AT} \propto \sum_{ \{ U_\ell = \pm 1 , V_\ell = \pm 1 \} } 
\prod_P \cosh \big[ \overtilde{\beta}(U_P + V_P) + \overtilde{\alpha} 
U_P V_P \big] \: .
\en
As the last step, we can rewrite this partition function in the usual
 Boltzmann form by defining
\eq
c\, e^{b(U_P+V_P)+a U_P V_P} = \cosh \big[ \overtilde{\beta}(U_P + V_P) + \overtilde{\alpha} U_P V_P \big] \: .
\en
where $a$, $b$, $c$ are suitable coefficients. Solving for $a$, $b$, $c$ we get
\begin{eqnarray}
	c^4    & = & \cosh(2\overtilde{\beta}+\overtilde{\alpha}) 
\cosh(2\overtilde{\beta}-\overtilde{\alpha}) \cosh^2 \overtilde{\alpha} \\
	e^{4b} & = & \frac{\cosh(2\overtilde{\beta}+\overtilde{\alpha})}
{\cosh(\overtilde{\beta}-\overtilde{\alpha})} \label{eq:sol-b} \\
	e^{4a} & = & \frac{\cosh(2\overtilde{\beta}+\overtilde{\alpha})
\cosh(2\overtilde{\beta}-\overtilde{\alpha})}{\cosh^2 \overtilde{\alpha}} 
\label{eq:sol-a}
\end{eqnarray}
therefore we can recast $Z_{AT}$ as the partition function of two coupled 
$\Z_2$ gauge systems
\eq
	Z_{AT}(\beta,\alpha) \propto \sum_{ \{ U_\ell = \pm 1 ,
 V_\ell = \pm 1 \} } \prod_P e^{b\,(U_P+V_P)+a\, U_P V_P} \,,
\en
where the duality transformation $\mathscr D : (\alpha, \beta) \to 
(a,b)$ from the AT couplings to the gauge couplings can be explicitly 
written, using (\ref{eq:sol-b}) and (\ref{eq:sol-a}), as
	\begin{eqnarray}
		a & = &
	 \frac{1}{4} \ln \Big( \frac{(\coth\beta+\tanh\beta\tanh\alpha)
(\coth\beta+\tanh\beta\coth\alpha)}
                    {2+\tanh\alpha+\coth\alpha} \Big) \\
		b & = &
	 \frac{1}{4} \ln \Big( \frac{1+\tanh^2\beta\tanh\alpha}{\tanh^2\beta+\tanh\alpha} \Big) \,\, .
	\end{eqnarray}
As a non-trivial check of these formulae one can verify that the $\mathscr{D}$ 
transformation is an involutory automorphism, i.~e.~$\mathscr{D}^2 = 1$, 
as required for any duality transformation.

\subsection{Dual of a Wilson loop}

The duality transformation maps any physical observable of the gauge theory  
into a corresponding observable of the spin model. In particular it is well 
known that the Wilson loops are related to suitable \emph{twists} of the 
couplings of the spin model. More specifically, let us consider a $\Z_N$ spin 
model in 3D. Let $\xi=e^{i 2 \pi / N}$ be the generator of this group. A 
$k$-twist of the link $\link{xy}$ in the spin action is defined by the 
substitution $\beta \to \xi^k \beta$ only in the selected link. 
Denoting with $Z_{\link{xy},k}$ the spin partition function modified 
in this way, one can easily prove the identity
\eq
	\avg{U_P^{(k)}}_{gauge} = Z_{\link{xy},k} / Z \,,
\en
where $P$ is the plaquette dual to $\link{xy}$ and $U_P^{(k)}$ is the 
plaquette variable in the irreducible representation of $\Z_N$ 
characterised by the integer $k = 1,2,\ldots,N$. A simple check is the 
following. The twist of the spin variable associated to a single node is 
equivalent to twisting all the links incident to this node; on the other
 hand the twist of a single spin can be re-absorbed in the invariance of the
 measure of the partition function. On the gauge side, this corresponds to
 the fact that the product of the plaquette variables lying on the six 
faces of the cube dual to the selected node is identically equal to 1.

Repeating the above construction for a suitable set of plaquettes we 
can construct in this way any Wilson loop or Polyakov-Polyakov correlator
 in any representation and its map into the spin model.

We want to fit this procedure to the AT model. In this case twisting a
 link corresponds to associating to it an anti-ferromagnetic coupling,
 i.~e.~ $\beta \to -\beta$. The character expansion (\ref{eq:char-1}) 
for an anti-ferromagnetic Boltzmann factor becomes simply
\eq
	e^{- \beta \,\sigma_x \sigma_y} = \sqrt{\frac{\sinh 2 \beta}2} 
\sum_{\mu_{\link{xy}}=\pm 1} \mu_{\link{xy}}\;
e^{\overtilde{\beta}\,\mu_{\link{xy}}}\; 
(\sigma_x \sigma_y)^{\,\hat{\mu}_{\link{xy}}} \,\, .
\en
Notice that in the AT model there are three different couplings associated 
to a single link. Which of them do we have to twist in order to build on
 the gauge side the plaquette in the fundamental (i.~e.~$k=1$) 
representation? To answer this question it suffices to twist, 
for instance, the variable $\sigma_x \to -\sigma_x$. From the point of 
view of the symmetries of the AT model, such a twist 
corresponds to a $\Z_4$ generator (\ref{genz4}) followed by a charge 
conjugation (\ref{chagen}); this corresponds to 
$\Psi_x\to-i\,\Psi_x^*$ in terms of the complex variable defined in (\ref{psi}),
hence it is associated to the fundamental representation. On 
the other hand this change of sign yields the twist of two couplings 
for each link incident to $x$: the quadratic $\sigma$ coupling and
 the quartic coupling. Therefore the plaquette in the fundamental 
representation is simply obtained by multiplying the Boltzmann factor 
by $\mu_{\link{xy}}\,\rho_{\link{xy}} = V_P$, where we used (\ref{eq:signum}). 
In this way, the identity outlined above can be written explicitly as
\eq
	\avg{V_P}_{gauge} = \avg{e^{-2(\beta+\alpha\tau_x\tau_y)
\sigma_x\sigma_y}}_{AT}
\en
generalising the known dual identity of the Ising model. Similarly, 
flipping the signs of both spins $\sigma_x$ and $\tau_x$ we get the
 plaquette variable in the $k=2$ representation as $\avg{U_P V_P}$. 
Combining together a suitable set of plaquettes we may build up any
 Wilson loop or Polyakov-Polyakov correlator with $k=1$ or $k=2$.

The reason why we insist in writing this model in terms of Ising 
variables is that here one can easily apply a very efficient non-local 
cluster method \cite{wd} which generalises in a straightforward way the
 one commonly used in Ising systems, based on the Fortuin-Kasteleyn (FK) 
cluster representation. Moreover, it has been built a very powerful 
method to estimate Wilson loops $\avg{W_\gamma}$ based on the linking 
properties of the FK clusters \cite{gv}: for each FK configuration 
generated by the above-mentioned algorithm one looks for paths in the
 cluster linked with the loop $\gamma$. If there is no path of this
 kind we put $W_\gamma = 1$, otherwise we set $W_\gamma=0$. 
This method leads to an estimate of $\avg{W_\gamma}$ with 
reduced variance with respect to the conventional numerical estimates.

\section{Monte Carlo simulations}

We performed a Monte Carlo analysis on the AT model, for both
 the fundamental and the double-fundamental strings, at the 
(confining) coupling $(\alpha,\beta)=(0.0070,0.1975)$ 
(for which we have measured the string tensions
 $a^2\sigma=0.01560(1)$ and $a^2\sigma_2= 0.0210(6)$ \cite{z4kstr_pietro}).

On the lattice, the quantity $P_\mathscr{R}(h)$ is measured by 
placing a square $R\times R$ loop $W(R)$ on a plane in the desired
 representation and taking the probe as a plaquette operator, 
parallel to the loop and lying on its axis at a distance $h$. 
The actual single measurement took into average also the four planar
 neighbours of the plaquette in the central position, in order to
 enhance the signal; this operation does not spoil the results 
since we dealt with large values of $R$.

According to the recipe for embedding the presence of a Wilson loop 
(in the representation $\mathscr{R}$) directly into the action as 
a series of frustrated links, the algorithm has only to measure the
 expectation value $\avg{U_p^{f}}^{W_\mathscr{R}}$ of the probe 
plaquette in the fundamental representation, where the superscript 
outside the average symbol denotes the fact that the action is 
modified to include the loop $W(C)$ according to (\ref{pfh})
\eq
	P_\mathscr{R}(h) \propto \avg{U_P^{f}}^{W_\mathscr{R}} +
 \mathrm{constant}\,.
\en

In terms of AT variables, measuring a plaquette operator translates to 
measuring the coupling energy on its dual link $\link{xy}$ (the sign 
variable $\epsilon$ is equal to $+1$ for every link except those dual to 
the loop, where its value reflects the applied frustration)
\eq
	\avg{U_P^{f}}^{W_\mathscr{R}}_{gauge} 
\stackrel{\sim}{\propto} \avg{\epsilon 
\sigma_x \sigma_y}_{AT} + \mathrm{constant}\,.
\label{eq:praticamente}
\en

The algorithm used for the analysis uses a cluster update method 
\cite{wd} basically similar to the standard Fortuin-Kasteleyn cluster 
technique: each update step is composed by an update of the $\sigma$ 
variables using the current values of the $\tau$ as a background 
(thus locally changing the coupling from $\beta$ to $\beta \pm \alpha$
 according to the value of $\tau_x \tau_y$ on the link $\link{xy}$),
 followed by an update of the $\tau$'s using the $\sigma$ values as background.

Note that since the large loop is automatically handled by the update 
procedure and the probe is a loop of side $1$ there is no need to
 implement expensive topological analysis on the configuration to
 get the measures, allowing us to reach a performance, with the lattices we used,
 of about 0.35 seconds per update/measurement step (as on a single 
\mbox{Intel {\scriptsize{\textregistered}}} Xeon 3.2 GHz 64-bit processor).
We used a cubic $L^3$ lattice with side $L=80$ (we found there are no 
finite size deviations up to $R \simeq L/2$) and measured $N$ times the 
plaquette operator as discussed. Configuration results have been then
 packed in groups of 25 in a binning fashion to estimate variances.
 We performed the measurements on loop sides $R=11,13,\ldots,41$ with a 
statistics of 220875 measures (with independent configurations  for 
each $R$) for the
 fundamental representation and 470275 for the $k=2$ case.

\section{Results}
The argument based on the properties of the minimal surfaces describing the 
world-sheet of the underlying confining string suggests  a 
logarithmic growth of the mean square width of the flux tube for both 
 the fundamental  and the double-fundamental ($k=2$) string. Also the 
numerical values of the two functions $w^2_{k}(R)$ at fixed $R$ should
 coincide, but this prediction is easily spoiled by artifacts of the 
short-range difference between a bound state of two strings and a 
fundamental string.

\FIGURE{
\psfrag{Transversal P(h) for k=2 and R=23}{$~$}
\psfrag{<link>}{$\avg{\epsilon\sigma_x \sigma_y}_{AT}$}
\includegraphics[width=10cm,angle=270]{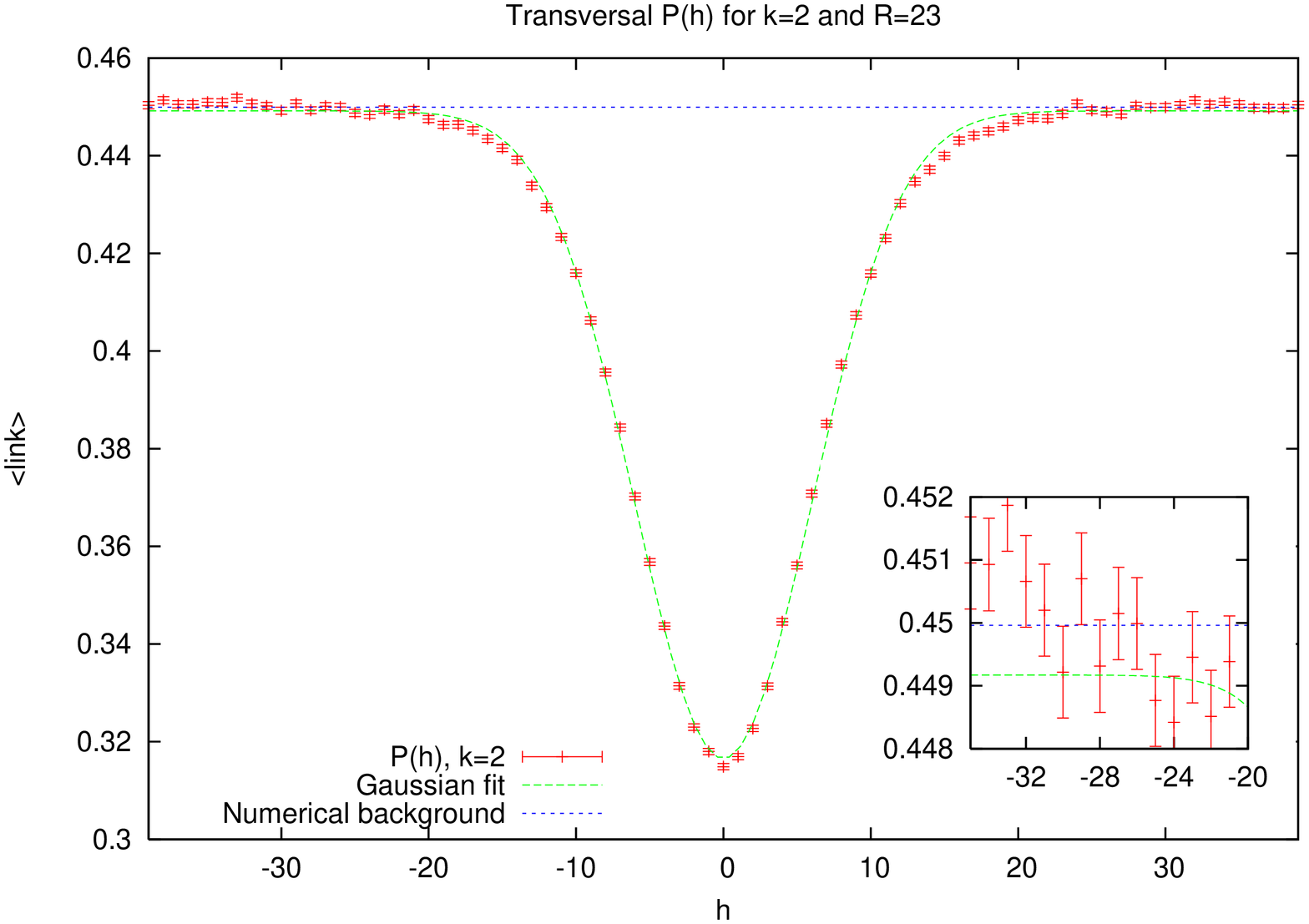}
\caption{Typical ($k=2$, $R=23$) transversal flux density distribution, 
as measured with the quantity in Eq.~(\ref{eq:praticamente}). 
Note the mismatch with the Gaussian form and the bad background
estimate that would follow from the Gaussian assumption for $P(h)$
\label{Figure:distribuzione}}}

\FIGURE{
\psfrag{R sqr(sigma)}{$R\sqrt{\sigma}$}
\psfrag{w^2 sigma}{$w^2\sigma$}
\psfrag{w2_FF}{w2 FF}
\psfrag{w2_F}{w2 F}
\psfrag{Flux tube width, F vs. FF}{$~$}
\includegraphics[width=10cm,angle=270]{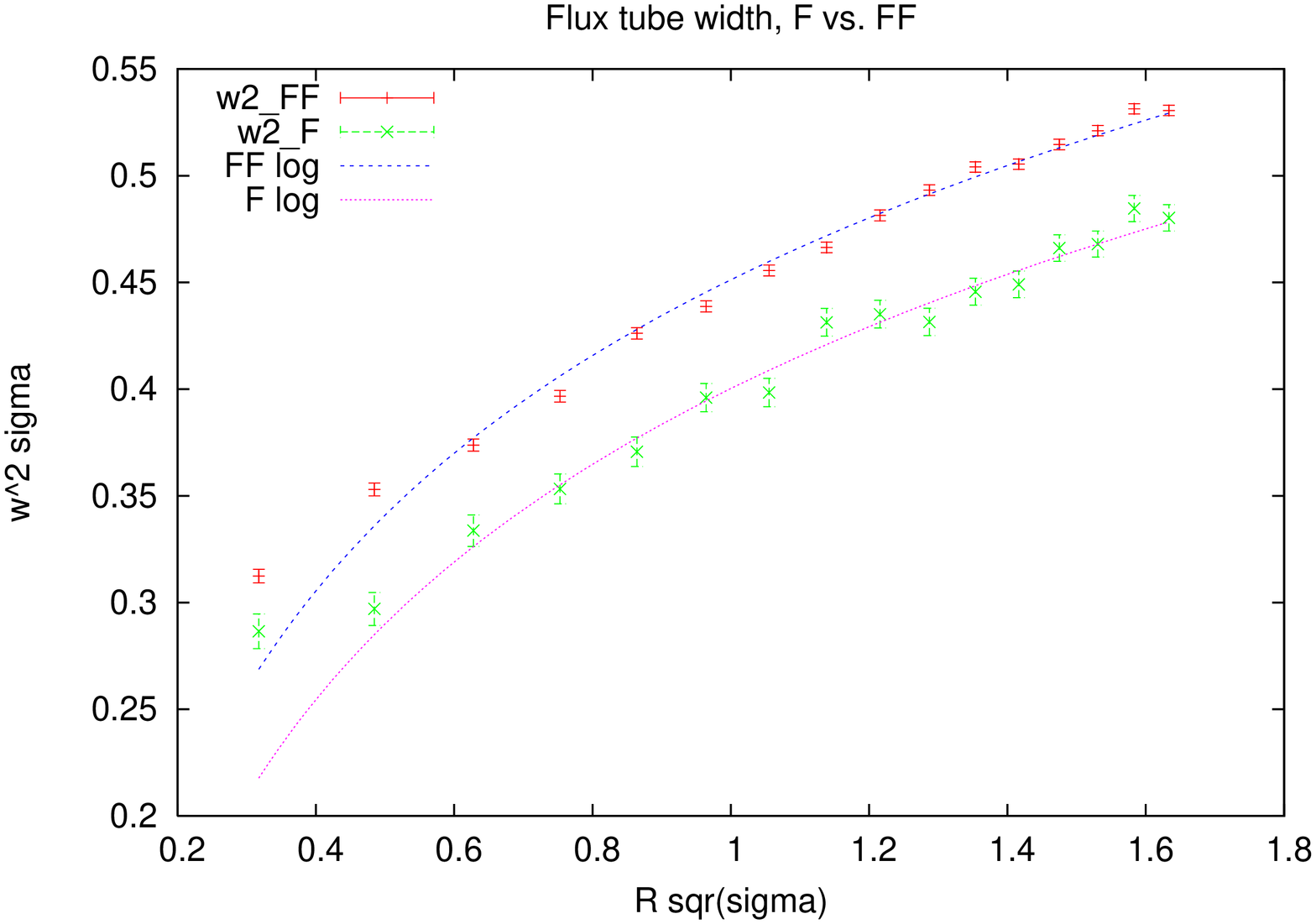}
\caption{Flux tube width, $k=1$ versus $k=2$\label{Figure:4}}}

We found that the measured values of $P_\mathscr{R}(h)$ do not
 fit too well to a normal distribution as approximately expected and 
lead to bad width estimates (see Fig.~\ref{Figure:distribuzione}),
 so we used the numerically integrated 
quantities instead, as in eq.~(\ref{eq:defw2}). We performed the
 integration by carefully choosing a cutoff value $h_{max}$: since the 
background value must be subtracted from the 
transverse flux density functions, the results are very sensitive to the 
choice of a cutoff. We took $h_{max}=12$.
To estimate this background value, we 
looked for a plateau in the region $|h|\geq h_{min}^{bg}$,
separately for each $R$, and took as final value the broadest averaged result in
 the range $h_{min}^{bg} = 18, \ldots, 24$.

By fitting the functions $w^2_{k=1,2}(R)$, for $R\geq R_{min}$ with $R_{min}$ 
an appropriate distance cutting off non-IR contributions, to the functional 
form
\eq
	\sigma\, w^2(R) = \frac{1}{2\pi}\log R + c
\en
(see Fig.\ref{Figure:4}) we found that the fundamental string width, 
as well as the 2-string, 
show the expected logarithmic growth with the appropriate universal
 multiplicative factor (reduced $\chi^2$ were, respectively, 1.22 and 2.68),
 but the value of $c$ differs measurably in the two cases: $c_1=0.4002(20)$, 
$c_2=0.4512(11)$; this discrepancy is probably due to some interaction
 between the fundamental world-sheets in the $k=2$ case.

Our work reinforces the numerical evidences of the predicted logarithmic
 broadening of the flux tube width \cite{cgmv}, extending them with high
 precision to the case of $\Z_4$, the simplest gauge group with more than one
 kind of k-string. 
In particular, the main result is that, when dealing with 
a non-fundamental string, its effective width still grows with the logarithmic
 law. This suggests that here the assumption of a free massless string behaviour
 for large $R$ holds from the $k=1$ case.
\section{Conclusions}
In the context of $\sun$ gauge theories we have studied the thickness of the 
confining flux tube generated by a pair of static sources in higher 
representations as probed by a Wilson loop in the fundamental representation, 
whose extension is small compared with the source separation. Generalising a
simple confining string picture proposed long time ago by 
L\"uscher M\"unster and Weisz \cite{lmw}  we argued that mean square width 
$w^2$, when measured in terms of fundamental string tension units $\sigma$,
grows logarithmically with the source separation $R$ in a manner which is 
universal, i.e.
\eq
\sigma\,w^2=\frac{D-2}{2\pi}\log\,R/R_c\,.
\en
The reference scale $R_c$  cannot be directly determined by the underlying 
string model and it is the only place where one can envisage a dependence on 
the source representation.

We performed a careful verification of these predictions in the case of a
3D $\Z_4$  gauge theory, which is the simplest gauge system where a 2-string 
forms. 

To reach the required high precision of the numerical data, the 
simulations where actually performed in the dual version of the system, 
which turns out to be a symmetric Ashkin-Teller model \cite{at}. 
This choice allowed us to use efficient 
non-local cluster algorithms \cite{wd}. The results of this analysis compare 
very favourably with the above predictions.

\end{document}